\documentclass[conference]{IEEEtran}
\IEEEoverridecommandlockouts
\usepackage{cite}
\usepackage{amsmath,amssymb,amsfonts}
\usepackage{algorithmic}
\usepackage{graphicx}
\usepackage{textcomp}
\usepackage{xcolor}
\usepackage{subcaption}

\def\BibTeX{{\rm B\kern-.05em{\sc i\kern-.025em b}\kern-.08em
    T\kern-.1667em\lower.7ex\hbox{E}\kern-.125emX}}
\begin{document}

\title{Maximizing net income of the auction waterfall with an abort decision tree}

\author{\IEEEauthorblockN{Michael Ting}
\IEEEauthorblockA{\textit{AlephD}\\
Paris, France \\
michael.ting@oath.com}
\and
\IEEEauthorblockN{Nicolas Grislain}
\IEEEauthorblockA{\textit{AlephD}\\
Paris, France \\
nicolas.grislain@oath.com}
}


\maketitle

\begin{abstract}
An online auction waterfall for an ad impression may contain auctions that are unlikely to result in a winning bid. 
Instead of always running through the full auction sequence, one could reduce the transaction cost by predicting and skipping these auctions. 
In this paper, we derive the auction abort rule that maximizes the net income of the waterfall under certain conditions, knowing only the publisher tag of the current auction and the ad request context.
The net income is defined as the payoff (revenue) minus the transaction cost. 
We translate the abort rule into a purity measure and propose a corresponding split criterion for a decision tree.
Training and testing on randomly sampled data indicate that the abort decision tree performs better than the full waterfall and the abort rule that makes use of only the publisher tag feature. When the transaction cost is higher, the cost saving, and thus net income gain, is higher for either abort decision rule.
\end{abstract}


\section{Introduction}
\begin{figure}[!ht]
\centering
\includegraphics[width=\linewidth]{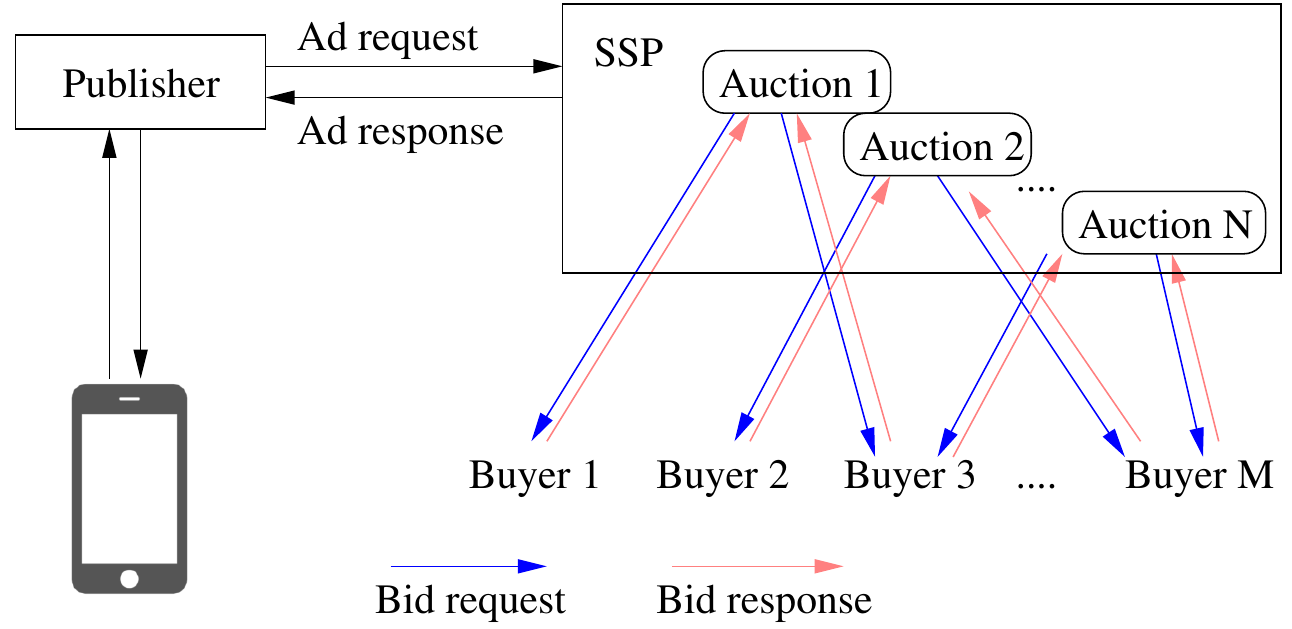}
\caption{Schema of an SSP conducting a waterfall auction for an ad impression.}
\label{fig:schema}
\end{figure}

An online ad supply-side platform (SSP) that seeks to maximize revenue from an ad impression has traditionally implemented an auction waterfall~\cite{pat:waterfall,web:aol_targeting}, where a sequence of auctions, each with different parameters, is held in order to sell the impression. 
The platform traverses the sequence of auction parameters, holding one auction after another, until a winning bid is found. 
When this occurs, the SSP stops the auction sequence and the impression is returned to the winning bidder. 
For example, the first auction could be intended for buyers with exclusive, first-look access to the inventory. More information could be disclosed on the impression, and the reserve price would be higher. As one goes down the auction waterfall, less information would be disclosed to the buyers, and the reserve price would decrease.
A schema of the process is shown in Fig.~\ref{fig:schema}.

The auction waterfall length for an ad impression can vary in length: a typical number would be from three to five. There are some ad impressions for which certain auctions in the waterfall are unlikely to produce a winning bid. Indeed, some ad impressions are not interesting for buyers and they may not even return bids in response to the bid requests. To simplify terminology, we shall also consider this to be a ``lost'' auction. 

A simplistic implementation of the auction waterfall would be to run through the entire auction sequence regardless of the quality of the ad impression. This is not optimal for the SSP, since there is a transaction cost to running auctions, e.g., network and machine costs. A more optimal approach is to only hold the auctions that are likely to succeed. More specifically, the SSP should hold an auction only if it has expected positive net income (NI), where the net income is defined as the difference between the payoff and the transaction cost.


Auction theory is a rich field, and has found applicability in various fields like online ads~\cite{a:stone_et_al,c:babaioff_et_al} and airline seat booking~\cite{a:subramanian_et_al}. 
While single-decision or static auctions were studied at the beginning~\cite{a:myerson}, there has been a trend towards studying dynamic auctions~\cite{a:vulcano_et_al,a:fang_et_al,tr:bergemann_said}. 
An advanced SSP would generate its auction waterfall dynamically to maximize revenue. 
To the authors' best knowledge, however, optimizing the auction waterfall to maximize \emph{net income} has not been considered in published literature before.

In this paper, we consider the case where the auction sequence is a priori unknown and seek, via a decision tree, to abort auctions that are unlikely to produce a positive NI for the ad impression. The abort decision is applied on an auction-by-auction basis.

\section{Problem description}
An ad impression opportunity generated by a user landing on publisher page creates an ad request $r$. This is resolved by holding a sequence of auctions in order to sell the impression. 
Suppose that there are $N$ auctions in the waterfall with publisher tags $t_1$, $\ldots$, $t_N$. The publisher tag is a unique identifier that defines the auction parameters. $N$ is not constant for all ad impressions, but is a configurable publisher setting. Thus, on an SSP, there are potentially waterfalls of different lengths for different publishers.
%
%

At the $i$-th auction, given $t_i$ and the ad request context $\theta$, we want to derive a decision rule on whether or not to abort the auction so as to maximize the expected net income of the ad request, defined as the expected payoff minus the expected transaction cost. 
Define $a_i \in \{0, 1\}$, $1 \le i \le N$ to be the abort flag for publisher tag $t_i$. If $a_i=1$, we abort the $i$-th auction; otherwise, we hold it.
If the $i$-th auction is aborted, the system continues to the next auction in the sequence and decides if it wants to hold that. Alternatively, if the $i$-th auction is carried out, the ad request is resolved if there is a successful bid; otherwise, the system proceeds to the next auction, just as in the case of $a_i=1$.

Let $f(r)$ be the payoff and $\zeta(r)$ be the number of \emph{played} auctions in the ad request $r$. Clearly, $1 \le \zeta(r) \le \min(N, |\{i : a_i=0\}|)$.
In order to simplify the problem, we \emph{assume} that there is a fixed cost $c$ to running an auction. This may not be true: for example, one could have a different number of bid requests for different publisher tags, and the network cost is proportional to the number of bid requests, c.f.~Fig.~\ref{fig:schema}. We want to therefore maximize the expected net income, which is
\begin{equation}
E[\textrm{NI}] = E[f(r)] - cE[\zeta(r)]
\label{eqn:net_income_criterion}
\end{equation}
where the expectation is taken of all ad requests $r$.
Suppose that $N$ is known. To maximize $E[\textrm{NI}]$, one would solve $\text{argmax}_{a_1,\ldots,a_N} E[\textrm{NI}]$ to obtain optimal $a_i$'s.

In our application, we do not know N a priori nor do we know the sequence of publisher tags. This lack of precise knowledge on the auction waterfall leads us to consider the problem of maximizing $E[\textrm{NI}]$ knowing only the ad request context $\theta$ and the publisher tag $t_i$ of the current auction.

\section{Derivation of the optimal abort decision rule to maximize the expected net income}
\label{sec:derivation_opt_abort_rule}

\subsection{Derivation}
\label{subsec:derivation}

Suppose that we are at auction $1 \le i \le N$ and want to decide if it is better to abort auction $i$ or not. 
%
Consider two cases.

\textbf{Case 1}. Abort auction $i$. \emph{Assume} that:
\begin{enumerate}
\item the expected payoff of the ad request $r$ is the \emph{same} as the case that auction $i$ loses, and
\item the expected number of auctions associated with the ad request $r$ is one less than in the case that auction $i$ loses
\end{enumerate}
The expected net income, taking into account the payoff and auction costs, is then
\begin{align}
& E[f(r)|\textrm{auction}_i\textrm{ loses}] - c(E[\zeta(r)|\textrm{auction}_i\textrm{ loses}] - 1) \nonumber \\
= & E[f(r)|t_i\textrm{ loses}] - c (E[\zeta(r)|t_i\textrm{ loses}] - 1)
\label{eqn:exp_net_income_a1}
\end{align}
where the conditional expectation is taken over all ad requests $r$ whose auction sequence includes the tag $t_i$, and $t_i$ loses.

Assumptions (1) and (2) may only approximately hold. For example, the auction sequence could be ordered so that the first few auctions maximize the revenue and the remaining auctions maximize the fill rate, i.e., the sale of the impression. 
Another possible scenario is that some buyers act strategically to withdraw or modify their bids in higher tiers of the waterfall since they anticipate being able to buy the impression more cheaply in a lower tier.
%
%

\textbf{Case 2}: Keep (hold) auction $i$. The expected net income is simply
\begin{equation}
E[f(r)] - c \cdot E[\zeta(r)]
\label{eqn:exp_net_income_a0}
\end{equation}
where the expectation is taken over all ad requests $r$ whose auction sequence includes the tag $t_i$.

The decision rule for $a_i$ is obtained by comparing the expected net income of the two cases, e.g., (\ref{eqn:exp_net_income_a1}) vs.~(\ref{eqn:exp_net_income_a0}). If $\textrm{Pr}(t_i\textrm{ wins}) = 0$, then $a_i=1$. Otherwise, if $\textrm{Pr}(t_i\textrm{ wins}) \neq 0$, after some simplification, one obtains
\begin{equation}
\begin{split}
& E[f(r)|t_i\textrm{ wins}] - E[f(r)|t_i\textrm{ loses}] \overset{a_i=0}{\underset{a_i=1}{\gtrless}} \\
& c\left(
	\frac{1}{\textrm{Pr}(t_i\textrm{ wins})} +
	E[\zeta(r)|t_i\textrm{ wins}] -
	E[\zeta(r)|t_i\textrm{ loses}]
\right)
\label{eqn:decision_rule}
\end{split}
\end{equation}
%

\subsection{Discussion}
\label{subsec:rule_discussion}

Consider some special cases in the application of the decision rule (\ref{eqn:decision_rule}).
\begin{itemize}
\item \emph{Zero transaction cost}. Then, $c=0$, and $a_i=0$ iff $E[f(r)|t_i\textrm{ wins}] > E[f(r)|t_i\textrm{ loses}]$. Irrespective of how small $\textrm{Pr}(t_i\textrm{ wins})$ is, it is still better to hold auction $i$.
\item \emph{Non-zero transaction cost and auction $i$ has a very small win probability}. The RHS of (\ref{eqn:decision_rule}) will be a large positive number and will most likely be higher than the LHS. In this case, it is better to abort auction $i$.
\end{itemize}

It is important to realize that the optimal decision rule given in (\ref{eqn:decision_rule}) maximizes the expected net income conditioned on knowing \emph{only} the publisher tag $t_i$. However, if other features are available, these should be used as additional conditioning variables as the expectations would be generally better estimators of the observed values. Consequently, one would expect a better abort decision rule.

\subsection{Adjustment for correlating effects}
\label{subsec:adjustment_correlating_effects}

The assumptions in Sec.~\ref{subsec:derivation} may not hold if there is a correlation between tag $t_i$ losing and $E[f(r)]$, $E[\zeta(r)]$.
Tag $t_i$ losing may be correlated with a lower expected payoff. Then, the expected payoff if the auction were aborted would be \emph{higher} than $E[f(r)|t_i\textrm{ loses}]$. 
There is an opposite effect on $\zeta$: tag $t_i$ losing may be correlated to a longer auction sequence. In this case, if the auction were aborted, the expected number of auctions would be \emph{smaller} than $E[\zeta(r)|t_i\textrm{ loses}]$.
The net effect is to decrease the LHS of (\ref{eqn:decision_rule}) and increase its RHS, making an abort decision more likely.

The critical terms in the decision rule are $E[f(r)|t_i\textrm{ wins}] - E[f(r)|t_i\textrm{ loses}]$ and $E[\zeta(r)|t_i\textrm{ wins}] - E[\zeta(r)|t_i\textrm{ loses}]$.
One way to reduce the correlating effect on the payoff is to reweigh observations in the calculation of the difference of expectations. 
Define $\tilde{b}(r)$ to be the median bid of ad request $r$. Instead of $E[f(r)|t_i\textrm{ wins}] - E[f(r)|t_i\textrm{ loses}]$, compute instead
\begin{equation}
E[E[f(r)|t_i\textrm{ wins}, \tilde{b}(r)] - E[f(r)|t_i\textrm{ loses}, \tilde{b}(r)]] 
\label{eqn:correct_correlating_effect}
\end{equation}
where the outer expectation is over $\tilde{b}(r)$ and set $\textrm{Pr}(\tilde{b}(r)|t_i\textrm{ wins})=\textrm{Pr}(\tilde{b}(r)|t_i\textrm{ loses})=\textrm{Pr}(\tilde{b}(r))$, i.e., the distribution of $\tilde{b}(r)$ is independent of $t_i$ winning or losing. Having identical median bid distributions in either a win or loss brings us closer, one hopes, to the first assumption in Sec.~\ref{subsec:derivation}.

To perform the computation, one must obtain a clustering of the bid medians $\tilde{b}(r)$. Let $B_1,\ldots,B_L$ denote a partition of the bid median support, where  $|B_j|\ge 1$ for all $j$. Then, (\ref{eqn:correct_correlating_effect}) can be calculated as
\begin{align}
\sum_{j=1}^{L} \{
& E[f(r)|t_i\textrm{ wins}, \tilde{b}(r) \in B_j] - \nonumber \\
& E[f(r)|t_i\textrm{ loses}, \tilde{b}(r) \in B_j] \} \; \textrm{Pr}(B_j)
\label{eqn:correct_correlating_effect_impl}
\end{align}
where $\textrm{Pr}(B_j) \approx$ the proportion of ad requests $r$ whose $\tilde{b}(r) \in B_j$.
A similar computation can be carried out for $E[\zeta(r)|t_i\textrm{ wins}] - E[\zeta(r)|t_i\textrm{ loses}]$. Here, we assume that $\tilde{b}(r)$ correlates to the number of auctions $\zeta(r)$.

\section{Design of the decision tree}

We noted in Sec.~\ref{subsec:rule_discussion} that additional features beyond the publisher tag should be used in the abort decision rule. For example, the ad request context $\theta$ includes user information, which is critically important for buyers of an ad impression. These user features would consequently also be predictive of whether an auction is worth running or not.

Decision trees are well known in the machine learning community~\cite{b:murphy}, having been used in different applications. 
%
%
%
We want to design an the abort decision tree classifier using  $\varphi := (\theta, t_i)$ as the available features in order to classify an auction as either ``abort'' or ``keep'', so $a_i=1$ and $a_i=0$ respectively.
Note that the payoff and auction waterfall length associated with each auction in the leaves are functions of the ad request $r$. The decision tree therefore incorporates auction sequence information.

A natural purity measure comes from the optimal decision rule (\ref{eqn:decision_rule}), where the expected net income between ``abort'' and ``keep'' is compared. We propose using the absolute difference of the expected net income (ADENI) when the auction is held vs.~when it is aborted. This is
\begin{equation}
\begin{split}
\textrm{AD}&\textrm{ENI} = | \textrm{Pr}(\textrm{auction wins}) \cdot \{ \\
& E[f(r)|\textrm{auction wins}] - E[f(r)|\textrm{auction loses}]) \\
- & c(E[\zeta(r)|\textrm{auction wins}] - E[\zeta(r)|\textrm{auction loses}])\} - c |
\label{eqn:impurity_measure}
\end{split}
\end{equation}
A larger ADENI is desirable. The split criterion is then the gain in the purity measure. Assuming an $n$-ary decision tree, and defining $\text{ADENI}(S)$ to be the ADENI computed over the set of observations $S$ one obtains
\begin{equation}
\textrm{ADENI}(S) - \sum_{j=1}^n p_j \textrm{ADENI}(S_j), \textrm{ where } p_j := \frac{|S_j|}{|S|}
\label{eqn:split_criterion}
\end{equation}
and $S_j,\; j=1,\ldots,n$ is a partition of $S$. Two stopping heuristics are used: when the size of a node falls below a threshold $T_\textrm{node}$, or when the increase in the ADENI purity criterion given by (\ref{eqn:split_criterion}) falls below a threshold $T_\textrm{ADENI}$.

ADENI can be adapted to (\ref{eqn:correct_correlating_effect}) by simply changing the way that the differences in expectations are calculated. 

\section{Experiments}

A binary decision tree ($n=2$) is estimated using the split criterion given by (\ref{eqn:split_criterion}), and is benchmarked against the simple decision rule (\ref{eqn:decision_rule}) that only takes into account the publisher tag.
Define $\textrm{NI}_{c,\eta}$ to be the net income assuming an auction cost of $c$ in CPM and the abort decision rule $\eta$, where $\eta \in \{0, s, t\}$ for when there is no abort rule, the simple rule, or the decision tree rule respectively. 
In order to calculate the performance of the decision rules $\eta \in \{s, t\}$, we use as the performance metrics the NI delta change $\Delta\textrm{NI}_{c,\eta} \overset{\textrm{def}}{=} \textrm{NI}_{c,\eta} - \textrm{NI}_{c,0}$ and the NI percent change $\%\textrm{NI}_{c,\eta}  \overset{\textrm{def}}{=} \Delta\textrm{NI}_{c,\eta} / \textrm{NI}_{c,0} \cdot 100$. 

Experimental data is drawn from 50 publishers on the Oath publisher platform.
%
%
The train dataset comprises randomly sampled data from one day's worth of auctions and the test dataset comprises randomly sampled data from the following day.
The train and test datasets have 195,122,188 auctions and 226,012,100 auctions respectively.
We consider three possible values for the auction cost $c$: \$0.003663, \$0.007326, and \$0.010989 in Cost Per Mille (CPM). The nominal $c=\$0.007326$ is the estimated transaction cost based on historical billing data, and we consider $c$ values $\pm 50\%$ around the nominal value. 
The test dataset's payoff is \$8,897.82. Using the nominal transaction cost, it has a net income of \$7,242.06.

The results are given in Table~\ref{tab1} below.
We tested out the adjustment in Sec.~\ref{subsec:adjustment_correlating_effects} to reduce the correlating effect of an auction loss to the payoff and auction sequence length. Since no significant difference was observed, these results are omitted for the sake of brevity. A possible explanation is that the assumptions in Case 1 of Sec.~\ref{subsec:derivation} approximately hold.
\begin{table}[!ht]
\centering
\caption{Performance comparison of the simple vs.~decision tree abort rule on the test dataset for different auction cost $c$ (in CPM cent).}\label{tab1}
\begin{tabular}{|l|l|l|}
\hline
Abort rule &  NI delta change $\Delta\textrm{NI}_{c,\eta}$ & NI percent change $\%\textrm{NI}_{c,\eta}$\\
\hline
\multicolumn{3}{|c|}{$c=0.3663$\textcent \;(CPM), $\textrm{NI}_{c,0}=$\$8,069.94}\\
\hline
Simple & \$42.06 & 0.52\% \\
Tree & \$102.72 & 1.27\% \\
\hline
\multicolumn{3}{|c|}{$c=0.7326$\textcent \;(CPM), $\textrm{NI}_{c,0}=$\$7,242.06}\\
\hline
Simple & \$187.31 & 2.59\% \\
Tree & \$355.71 & 4.91\% \\
\hline
\multicolumn{3}{|c|}{$c=1.0989$\textcent \;(CPM), $\textrm{NI}_{c,0}=$\$6,414.17}\\
\hline
Simple & \$404.36 & 6.30\% \\
Tree & \$820.87 & 12.80\% \\
\hline
\end{tabular}
\end{table}
The decision tree rule has better performance than the simple rule. As the auction cost $c$ increases, the NI percent change $\%\textrm{NI}_{c,\eta}$ increases for both abort rules $\eta$. This is because, when $c$ increases, its impact on the NI becomes more significant. If one is able to skip unsuccessful auctions, there is a noticeable savings gain to be realized.

\begin{figure}[!ht]
\centering
\begin{subfigure}[b]{.45\linewidth}
\includegraphics[width=\linewidth]{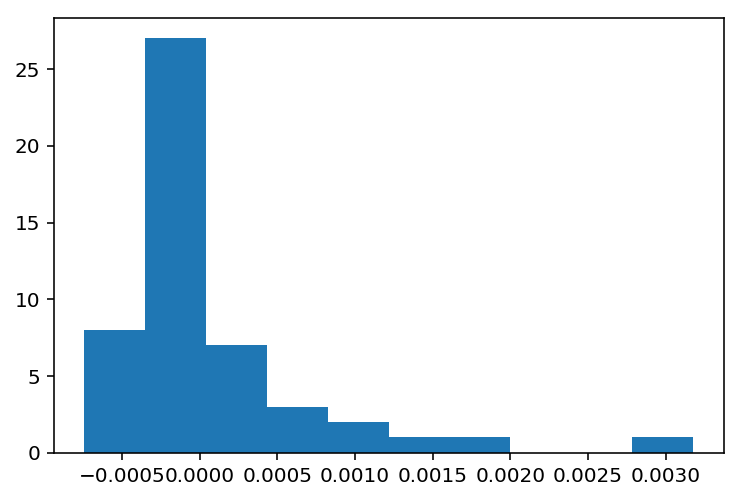}
\caption{$c=0.3663$\textcent, simple rule}
\label{fig:halfcost_simple}
\end{subfigure}
\begin{subfigure}[b]{.45\linewidth}
\includegraphics[width=\linewidth]{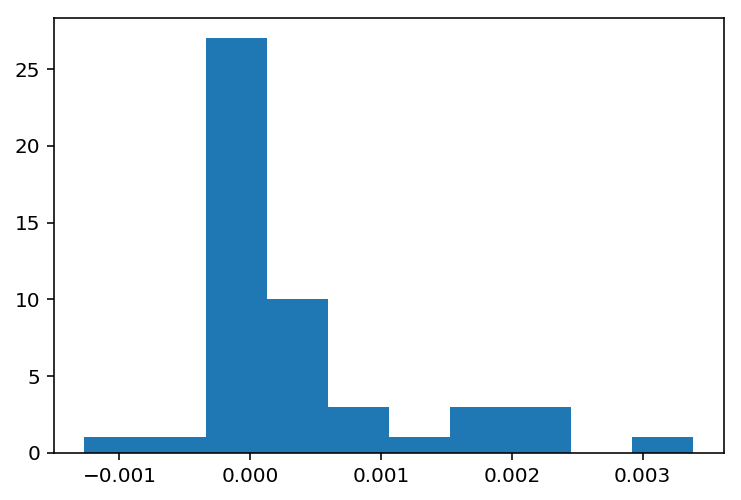}
\caption{$c=0.3663$\textcent, tree rule}
\label{fig:halfcost_tree}
\end{subfigure}
\begin{subfigure}[b]{.45\linewidth}
\includegraphics[width=\linewidth]{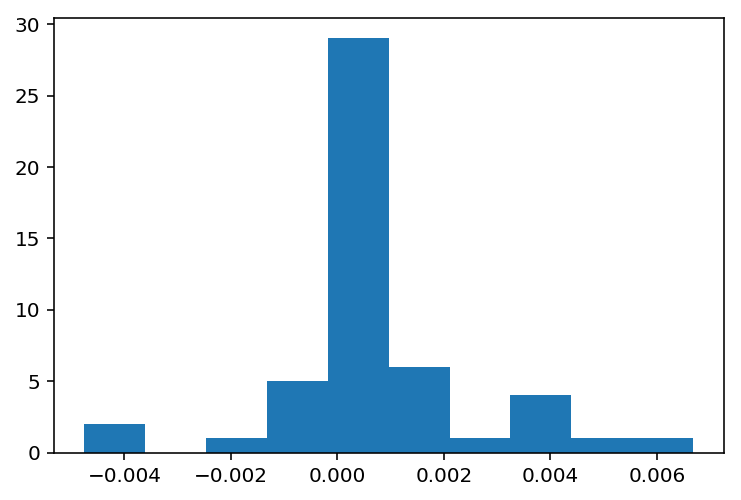}
\caption{$c=0.7326$\textcent, simple rule}
\label{fig:cost0_simple}
\end{subfigure}
\begin{subfigure}[b]{.45\linewidth}
\includegraphics[width=\linewidth]{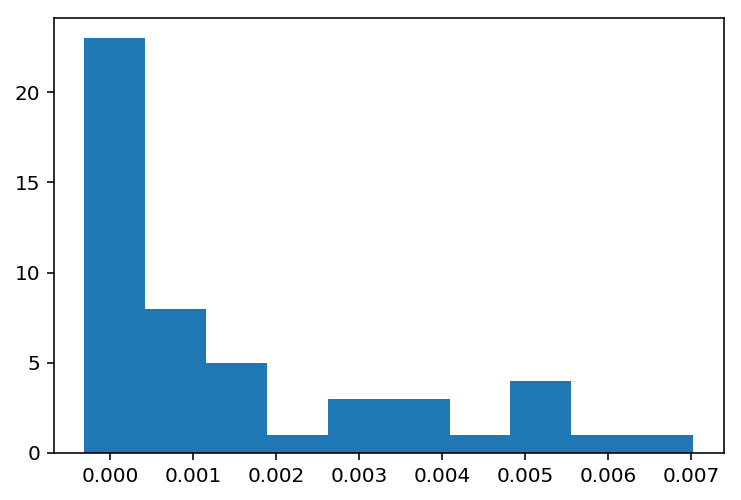}
\caption{$c=0.7326$\textcent, tree rule}
\label{fig:cost0_tree}
\end{subfigure}
\begin{subfigure}[b]{.45\linewidth}
\includegraphics[width=\linewidth]{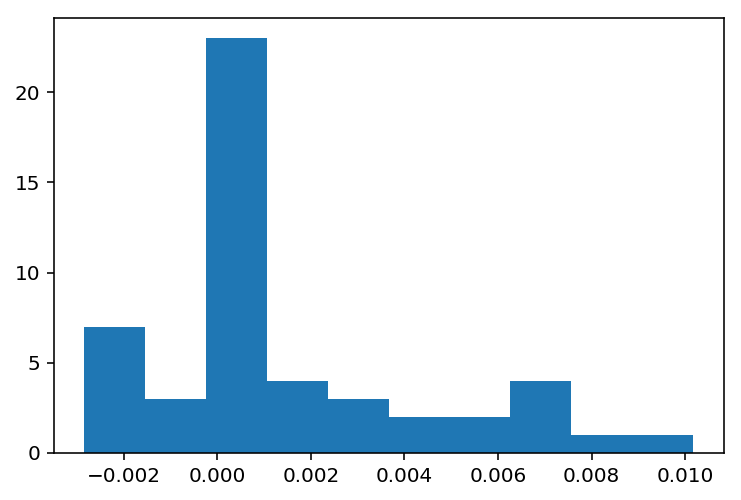}
\caption{$c=1.0989$\textcent, simple rule}
\end{subfigure}
\begin{subfigure}[b]{.45\linewidth}
\includegraphics[width=\linewidth]{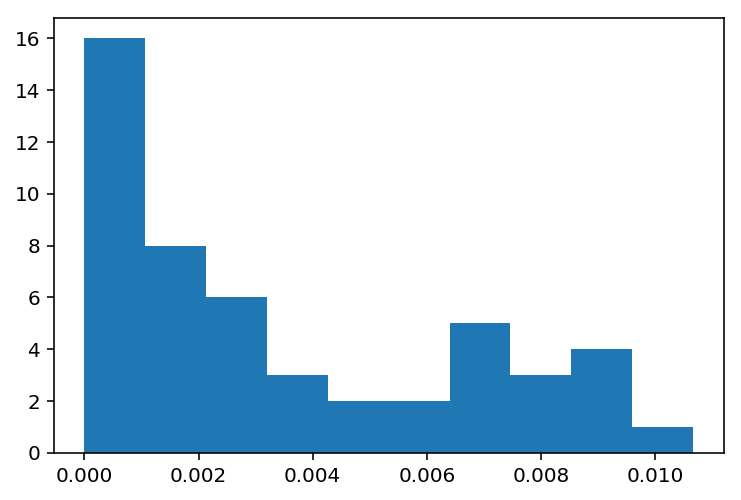}
\caption{$c=1.0989$\textcent, tree rule}
\end{subfigure}
%
\caption{Histograms of average NI delta change in CPM dollar at different auction costs $c$ (in CPM cent) for the simple and decision tree-based abort auction rules.}
\label{fig:histogram_experiment}
\end{figure}

The savings gain varies from publisher to publisher. In Fig.~\ref{fig:histogram_experiment}, we plot histograms of the average NI delta in CPM dollar for each publisher at different values of the auction cost $c$.
The average NI delta is defined as the delta NI divided by the number of ad requests. From the histograms, this number generally increases in magnitude as the auction cost $c$ increases. The benefit of aborting an unprofitable auction appears to increase the higher the auction cost.
Irrespective of $c$, the simple rule always has several negative average delta NI. The decision tree rule, which makes use of user-specific features, has non-negative performance for all publishers when $c=0.7326$\textcent \; and $c=1.0989$\textcent.  When $c=0.3663$\textcent, there are several publishers for which the average delta NI is negative.

\section{Conclusions and future work}

Under certain conditions, we derived the optimal auction abort decision rule to maximize the expected net income given only the publisher tag of the current auction and the ad request context. A purity measure and splitting criterion were proposed to construct an abort decision tree.
%
%
In the experiment conducted, the decision tree rule that takes into account additional features like user-related features performs better than the simple rule that just uses the publisher tag feature.
In order to confirm the results of the experiments, we would have to A/B test. If we find that the explore cost opportunity is significant, we could consider contextual bandits.

%
A better abort decision rule could be achieved if more information were available regarding the auction sequence, e.g., if the auction sequence were known a priori, or if auction sequence probabilities were known or could be estimated. Indeed, the decision rule and decision tree derived in this work did not even make use of the auction index $i$.
%
%
To illustrate how much information is contained in additional knowledge of the waterfall, suppose that the SSP is at publisher tag $t_i$ but knows that tag $t_{i+1}$ has a higher expected payoff. It makes sense then to skip the current auction and move on to the next one in the sequence. 
%
A limited form of waterfall optimization has been explored in this work in the sense that, rather than start out with a sub-optimal waterfall designed to maximize revenue and prune off unprofitable auctions, one could design the auction sequence to maximize net income from the very start.

\bibliographystyle{IEEEtran}
\bibliography{IEEEabrv,mt}


\end{document}